# Collaboration Conundrum: Synchrony-Cooperation Trade-off

Tamas David-Barrett

Email: tamas.david-barrett@trinity.ox.ac.uk

Address: Trinity College, Broad Street, Oxford, OX1 3BH, UK

Web: www.tamasdavidbarrett.com

**Abstract**

In large groups, every collaborative act requires balancing two pressures: the need to achieve behavioural synchrony and the need to keep free riding to a minimum. This paper introduces a model of collaboration that requires both synchronisation on a social network and costly cooperation. The results show that coordination slows, and cooperativeness increases with the social network's local integratedness, measured by the clustering coefficient. That is, in a large-group collaboration, achieving behavioural synchrony and strategic cooperation are in opposition to each other. The optimal clustering coefficient has no natural state in our species, and is determined by the ecological environment, the group's technology set, and the group's size. This opens the space for social technologies that solve this optimisation problem by generating optimal social network structures.

# Introduction

Achieving a collaborative act in a large group of people comes with a difficulty because it requires balancing two pressures: the need for achieving behavioural synchrony and the need to keep free riding to a minimum (1-3). Our gut instinct might suggest that these two pressures could be automatically aligned: in a healthy community we could expect that both coordinating a collective task, as well as the lack of shirking would go hand in hand. However, this insight compares 'healthier' to 'less healthy' communities, whatever these terms mean, rather than specify the characteristics of the social network of a particular group of people.

Let us define collaboration as the problem a set of agents face when agreeing to form a group that will perform a collective action that requires both coordination and costly individual cooperation. Efficient collaboration, thus, needs solutions for two different kinds of collective action problems.

First, any such act requires the synchronisation of what the action will be, a plan of individual participation, and the coordination of the individual actions towards the collective act. That is, one necessary condition for a collaborative act to take place is that the group achieves behavioural synchrony (4). The agents need to come to an agreement on what they want to achieve, who will do what, and how what they do will make sense together (5-7).

Second, if the individual action is costly, collaboration in a group carries the risk of free riding (8-15). In most forms of complex collective action, the optimal action by an individual is costly to the individual, while beneficial for the group: a space for free riding emerges (16, 17). When this happens, the collaboration faces a second problem: it is essential to reduce the incentive for the individuals to shirk from their assigned, agreed-upon individual tasks.

Mutualistic collective action is a counterexample that illustrates the point. Mutualism requires only that agents do what they would do anyway. For instance, predator avoidance is mutualistic cooperation (18), in which there is no strategic element: the agents keep together, forming a herd of ungulates (19-21), or a school of fish (22, 23), or a flock of birds (24, 25). They do so because the individual action reduces the individual probability of predation (26, 27).

Thus, collaboration raises two parallel problems: that of achieving behavioural synchrony and that of free riding (28).

In some cases, the two problems are independent from each other. For instance, when the free riding problem is solved dyadically, e.g., in reciprocal 'altruism' (29-32), the synchronisation problem's solution can ignore the problem of free riding.

However, many forms of collective action take place coordinated on large social networks, in which the dyadic reciprocity-based guarantee of agents pulling their weight is too fragile. In these, inclusive fitness-based solution may emerge (33-35), which could also be independent of the synchronisation problem. Furthermore, the solutions to the two problems could be supporting each other, as is the case of the eusocial insects that are able to keep free riding to a minimum while achieving behavioural synchrony in very large groups (36-42).

Human collective action tends to emerge on social networks which are too large to be maintained solely based on relatedness, a characteristic that our species shares with all other apes. Thus, we have a problem: cooperative stance of collective action cannot be maintained via dyadic reciprocity and relatedness alone. To solve this problem, humans, in all cultures, form social networks in which cooperativeness is tracked using third-party information (43). The ability to use such reputation-based tracking of cheaters in a social network, however, is dependent on structural properties of the graph (43-65). Different logics of organising a social network end up in different structural properties (66, 67). While these solutions all result in increased cooperativeness, it is not clear that they are at the same time not harming the group's ability to coordinate the collaborative act.

It has already been shown that a social network's average clustering coefficient can be associated with increased cooperative stance (8-11, 43). However, high clustering coefficient can have a negative impact on macro-level processes (1). For instance, when status-based stratification alters a social network's structure, and thus increases the clustering coefficient, reaching behavioural synchrony among the agents becomes slower (68).

This paper's objective is to investigate the effect of a social network's integratedness, measured by the average clustering coefficient, on the two constituent elements of collaboration: that of coordinating the collective action and that of reducing free riding. In this, this paper aims to bring together two strands of the literature: behavioural synchrony and origins of cooperation.

# Methods and Results

To measure the effect of the integratedness on cooperation and synchronisation, I needed a sample of graphs that covers the entire range of clustering coefficient.

For this I used a recently built graph library (69). This library contains an unbiased sample of connected 4-regular graphs of the size $n$=10 to 50 in increments of 5. For each $n$, the sample contains 100 non-isomorphic graphs for each possible clustering coefficient value. These sub-samples are unbiased in the basic graph properties. NB. The clustering coefficient's range of $k$=4 is 0 to 0.7. For the cases when the clustering coefficient is near the maximum, fewer than 100 graph structures exist. For these bins, the sample contains less than 100 graphs, and these are not entirely unbiased. (See details in 69.)

For each graph, I ran both (i) a network reputation game (following 43), and (ii) a synchronisation algorithm (following 4). From these I captured the effect of the clustering coefficient on the two respective problems: (i) the number of repeats needed to maintain cooperation, and (ii) the efficiency in achieving behavioural synchrony.

### Network reputation game

The following game is identical to the setup in (43), but ran on a new, standardised graph library and a wider range of graph sizes (69).

Let us define a social group made of $n$ agents who form a social network, $g$ such that the degree is uniform and fixed at $k$. The agents have two forms of interaction along the edges: (i) they can play a game with each other, and (ii) they can pass on information about third parties.

The agents play dyadic prisoner's dilemma games where the payoff matrix is symmetric, and is symmetric and ((1, -1.6), (1.5, 0)). Thus payoffs are set so that the Nash Equilibrium is cheat-cheat. (For a discussion about the particular values, see 43.)

Let us assume that one player is randomly chosen as having type 'cheater', and thus $n$-1 players have type 'cooperator'. The players play the game in rounds. For each round, two connected agents, $i$ and $j$, are randomly selected.

The players track others' expected types the following way. Let $d_{i,j}$ denote agent $i$'s knowledge of $j$'s type, with the initial value of co-operator for all. Let $e_{i,j}$ denote the private expectations held by agent $i$ about agent $j$'s type. The updating of these expectations is done the following way:

$$e_{i,j} = \begin{cases} cheater \text{ if } d_{i,j} = cheater \\ cheater \text{ if } d_{i,j} = cooperator \text{ and } \dfrac{\tilde{k}}{k+1} \geq 0.5 \\ cooperator \text{ otherwise} \end{cases}$$

where $\tilde{k}$ is the number of trusted friends who think that $j$ is a cheater, and trusted friends are those agents that are connected to her, and she still thinks that their type is cooperator.

Let us set the action of $i$ based on this expectation:

$$a_i = \begin{cases} cheat \text{ if } e_{i,j} = cheater \\ type_i \text{ if } e_{i,j} = cooperator \end{cases}$$

The two players, $i$ and $j$, move at the same time, collect their payoffs, and update their knowledge about each other's types:

$$d_{i,j} = a_j$$
$$d_{j,i} = a_i$$

Let the interactions repeat until an average agent is in play $r$ times, at which point, I recorded the cheater payoff and the average of non-cheater payoffs.

This algorithm was used on each of the 100 graphs in each of the clustering coefficient bins in the graph library, with network size going from 10 to 50 in increments of 5.

Representing the difference between the cheater and non-cheater payoffs as a function of the repeat number (Fig. S1.) allows the calculation of the minimum number of repeats required for the maintenance of cooperation, i.e., the point at which the cheater and the non-cheater payoffs are equal, denoted as $r^*$. I calculated these points by linear interpolation of the datapoints.

The results show that the clustering coefficient drives $r^*$: increasing clustering coefficient makes the maintenance of cooperation easier for all $n$ (Fig. 1).

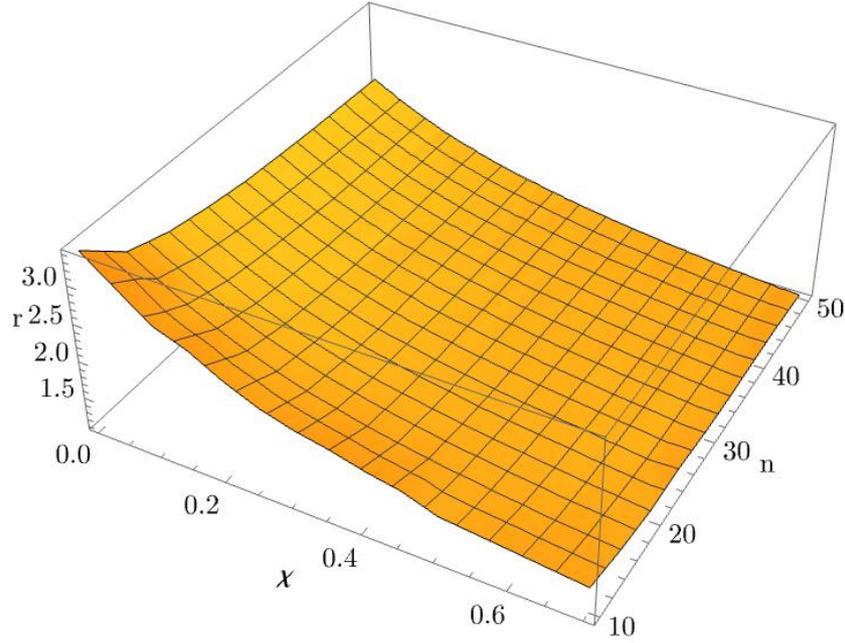

Fig. 1. The effect of the clustering coefficient on the number of repeats needed for cooperation to emerge on a graph.

### Measuring the efficiency of group coordination

To measure the speed of coordination, I used the behavioural synchrony model (4, 70), in which coordination takes place on a unit circle, which is a proxy for finding a shared compass direction (71-73).

Let $n$ agents form a connected graph of degree $k$, as is the case of all elements of the graph libraries generated above. Every agent is randomly assigned an initial value drawn from a uniform distributed variable on the unit circle:

$$\phi_{0,i} \sim U(0°, 360°)$$

Connected pairs of agents are randomly selected, and synchronise their compass directions by setting their $\phi$ values to the mid-point of their old $\phi$ values:

$$i, j \sim U\{1, , n\} | i \neq j$$

$$\phi_{t+1,i} = \begin{cases} f1 & \text{if } 0° \leq f1 \leq 360° \\ f1 + 360° & \text{if } 0° > f1 \\ f1 - 360° & \text{if } 360° < f1 \end{cases}$$

where

$$f1 = \begin{cases} \dfrac{\phi_{t,i} + \phi_{t,j}}{2} & \text{if } |\phi_{t,i} - \phi_{t,j}| \leq 180° \\ \dfrac{\phi_{t,i} + \phi_{t,j} - 360°}{2} & \text{if } \phi_{t,i} - \phi_{t,j} > 180° \\ \dfrac{\phi_{t,i} + \phi_{t,j} + 360°}{2} & \text{if } 180° < \phi_{t,i} - \phi_{t,j} \end{cases}$$

Let $\delta$ denote the degree of compass deviation among the agents:

$$\delta_t = \sum_i \sum_j |\phi_{t,i} - \phi_{t,j}|$$

where $t$ is the number of meetings an average agent has.

The results show that (a) increasing clustering coefficient, $\chi$, reduces coordination efficiency, i.e., $\delta$ increases, and (b) this effect is present for all $n$ (Fig. 2).

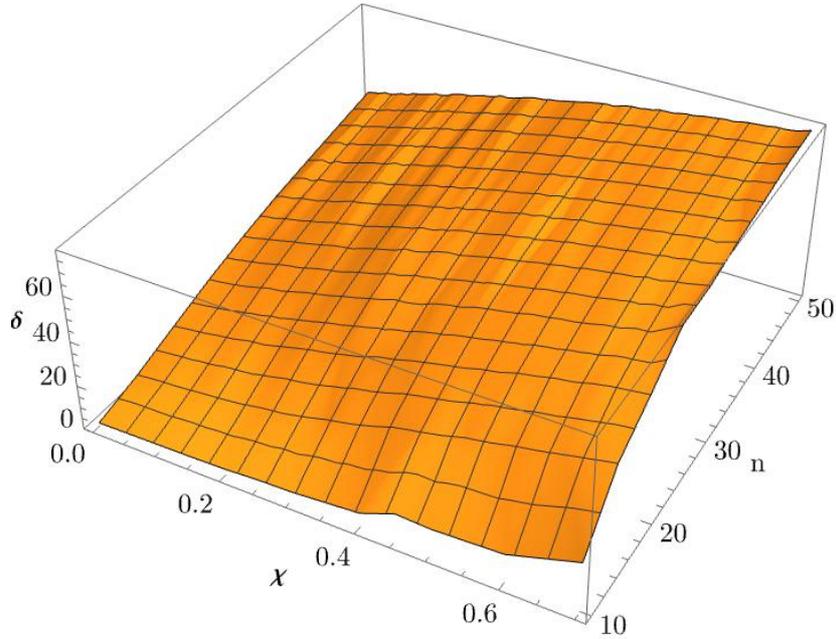

Fig. 2. The effect of the clustering coefficient on the speed of achieving behavioural synchrony. (Notice that the z-axis is inverse delta, thus a form of measuring of coordination speed.)

## Synchrony-cooperativeness trade-off

Notice that the number of repeated interactions that are necessary for maintaining cooperation, and the synchronisation rounds that are necessary for being able reach behavioural synchrony, can be regarded as two different types of costs of collaboration. The results above suggests that they have opposing relationships with the microstructure of the social network: high clustering coefficient yields higher cooperativeness (Fig. 1), but slower synchronisation (Fig. 2).

Let us normalise the two measures ($r^*$ and $\delta$), so that their magnitudes are comparable. The result illustrates the trade-off between these two constituent elements of cooperation: the cost of reputation-based cooperative stance decreases with the interconnectedness of the network, while at the same time, achieving behavioural synchrony slows down (Fig. 3).

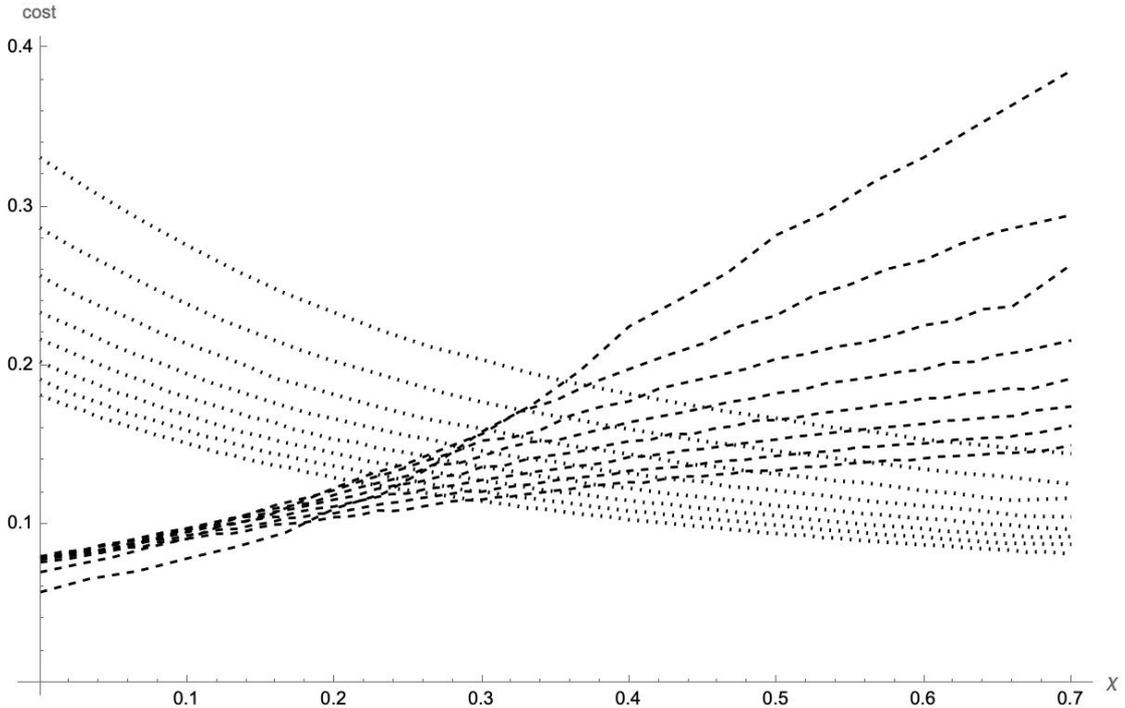

Fig. 3. The trade-off between synchronisation and cooperation is due to their inverse relationship with the clustering coefficient. Each line represents a normalised cost, either of synchronisation (dashed lines) or maintaining cooperation (dotted lines), where the group size is increasing across both batches of lines (counting on the right-hand side); x-axis: clustering coefficient; y-axis: normalised $\delta$ and $r$, respectively.

## Optimal clustering coefficient

As an illustrative example, let us introduce a combined cost function of the arbitrary functional form: cost=$r^3+\delta^3$, such that both factors are normalised between 0 and 1 for each $n$ bin.

Fig. 4 depicts the result. The cost minima are at the mid-range clustering coefficients in the case of this illustrative cost function.

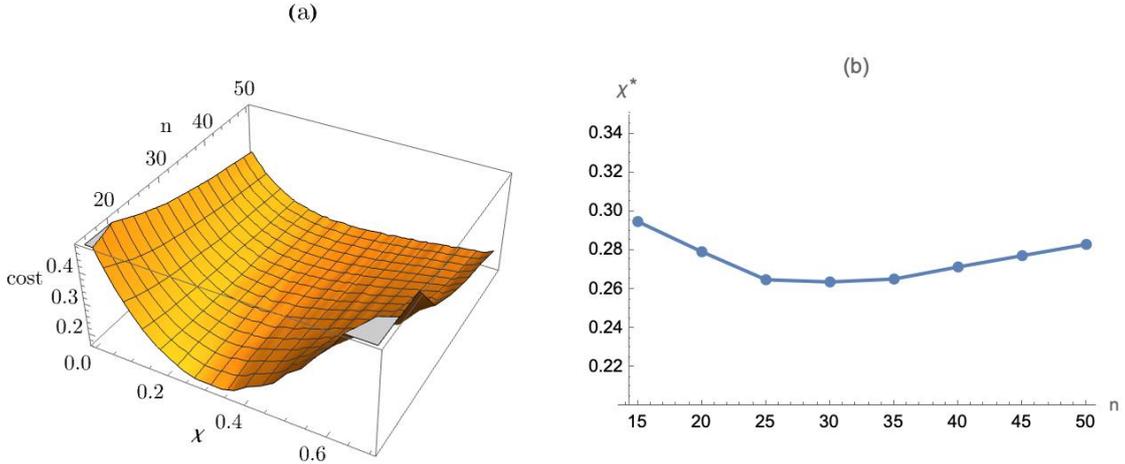

Fig. 4. Optimal clustering coefficient can be a meaningful concept and is likely to be mid-range when both factors are important. Panel (a): the cost minimising clustering coefficient is at mid-range for all group sizes for the cost function example; panel (b): cost minimising optimal clustering coefficient as a function of group size, $n$.

## Two further observations

Observation 1: In Fig. 4, both the cooperation threshold and the speed of coordination are normalised within each $n$-bin. This allowed the observation that the optimal clustering coefficient is likely not to be a corner solution, independent of group size. However, normalising within $n$-bins, rather than across, is misleading if our question concerns the effect of group size on the optimal clustering coefficient. Normalising across all bins, and using the same (illustrative) cost function, changes the result: as the group size increases, the clustering coefficient needs to fall to allow the coordination to take place at an acceptable speed (Fig. 5). NB. This modification is equivalent to giving increasing weight to synchronisation efficiency when group size is increasing.

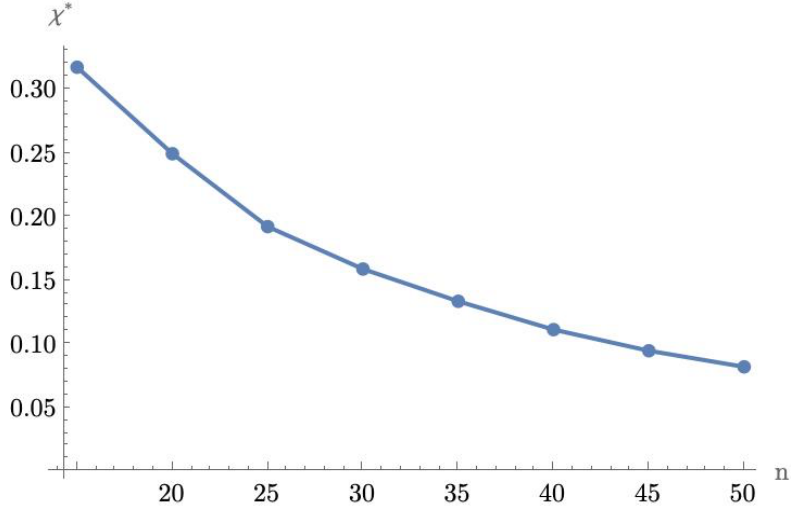

Fig. 5. When the costs are comparable across the different n-samples, increasing group size leads to falling optimal clustering coefficient.

There is an alternative way of defining the cost of synchronisation: not via synchronisation efficiency, but rather the number of meetings it takes to reach a particular, fixed level of synchrony. That is, making synchronisation speed, $t$, be dependent on a fixed $\delta$, rather than the other way around. The model of this paper does not use this measurement, because for high clustering coefficient cases it can take a forbiddingly long computation to reach synchrony. If there is a limit to how many interactions an agent can engage in, and if $\chi$ is high then in large groups the coordination effectively halts. They cannot get there before the interaction number limit is reached. This observation is in line with the suggestion that increasing group size should be associated with higher weight to synchronisation efficiency, when the cost is defined as in this paper's model. Thus, it is also in line with the qualitative result of Fig. 5, that is, with increasing group size, the optimal clustering coefficient falls, and thus also the group's general cooperativeness.

Notice that with the fall in the clustering coefficient, cooperativeness is enforced by building dyadic bonds, and the use of network reputation mechanics for maintaining cooperativeness becomes less frequent. This makes cooperation overall more expensive and opens the possibility for alternative solutions for maintaining cooperative stance, for instance by payoff-altering institutions like legal systems. This observation is in line with Ostrom's non-mathematical insight that only relatively small groups are likely to sort out free riding among themselves (74, 75).

Observation 2: Although, by definition, both the micro-structure's integratedness and the graph's size are fixed for each $\chi$-$n$ bin, there is space for meso-structure variation among the elements of these sub-samples. After all, for n>10, the number of non-isomorphic graphs is extremely high, e.g., >10^11 for n=20, and >10^49 for n=50 (69). Indeed, this structural variation leads to a variation of the costs associated achieving behavioural synchrony while maintaining a network wide cooperative stance. Using the illustrative cost function further up, Fig. 6 shows substantial structural diversity among the sample elements.

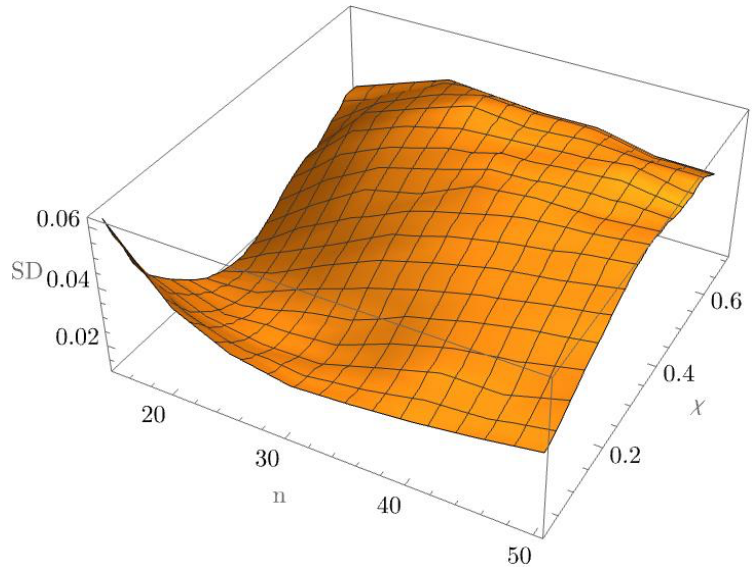

Fig. 6. Cost variation inside each $\chi$-$n$ bin. (Bin-wide standard deviation of the c=$r^3+\delta^3$.)

This observation allows us to calculate the relationship between cost and group size. Fig. 7 depicts the maximum $n$ that can be achieved at different cost levels. Not surprisingly, the higher the cost, the higher is the maximum group size. However, there is substantial gain in group size when using each $\chi$-$n$ bin's best performing graph compared to the mean-field.

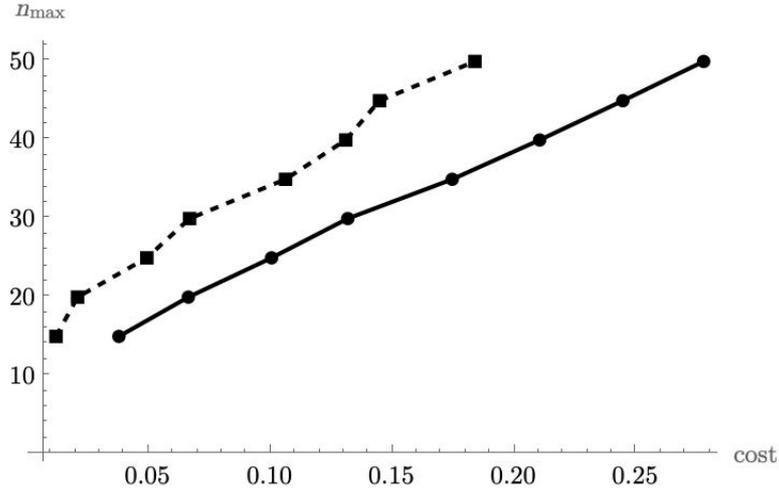

Fig. 7. Maximum group size as a function of cost. Continuous line: mean field, dashed line: best in the sample.

That is, allowing the graph meso-structure to vary allows larger groups with the same resource cost.

# Discussions

This paper's results suggests that in large groups collaboration faces two different and opposing pressures: cooperativeness increases as the network is more integrated, but synchronisation speed falls at the same time. This result has structural consequences for the social network.

### Consequence: optimal clustering coefficient

(A) The optimal clustering coefficient is likely to be in-between if both factors are important, that is, neither maximum nor zero, with the cost function determining the particular optimum point.

(B) If cooperativeness is essential, but synchronisation is not, the optimal clustering coefficient is likely to be high. However, in this case, it is not clear why there is a group to start with. If there is nothing to coordinate, then to what extent could the group be regarded as a meaningful category, rather than a set of loosely connected smaller communities? Perhaps, there are cases when there is a biological need for such a group (e.g., for the purpose of maintaining genetic diversity), but in these cases, the relevant concept is 'population' rather than 'group', and hence the graph is that of a population structure.

(C) If coordination is essential, but cooperativeness is not, then the optimal clustering coefficient is likely to be zero. This is essentially a herd.

In modern human terms, an example for (A) could be a university faculty, for (B) people working in small firms on a competitive market, in which case the market provides the synchronisation mechanism, and for (C) commuters going to work.

### Consequence: social technologies

Notice that there is no 'naturally' set clustering coefficient for a human group. As far as we know, there is no inherited regulation of behaviour that would end up in one or other structure of our communities.

In other words, there is no reason to think that any in-between clustering coefficient level would emerge automatically. As the examples in this paper showed, the optimal integratedness depends on the shape of the cost function and the size of the group, which, in turn, are determined by the technology-mediated effective ecology in which the group exists. That is, not only there is no natural state of the social network structure, but its optimum shifts either as the ecology or the technology that exploits shift, or as the population size changes. This yields a space for social technologies that could perform this function of structural regulation.

Let us consider three types of social technologies: (i) one that sets an in-between clustering coefficient, (ii) one that eliminates the cooperation problem by solving it through other means, and (iii) one that eliminates the coordination problem.

(i) To find an example for a social technology that sets an in-between clustering coefficient, we need to think of a cultural regulatory system that facilitates the rise of a space for reputation-based cooperation through locally integrated network structure, but also limits the space for delineated cliques. Modern examples could be the settlement planning of suburban towns, or corporate working space design of open offices, e.g., in investment banking trading floors. On a trading floor there are many different teams in which team members rely on each other, in a costly cooperation setup, but also cooperate with some other team members across the floor. The design of the floor, i.e., who sits where, is important because if the teams become cliques, the coordination across the floor collapses, for instance, information flow about a looming crisis is impeded, while if the team members are not interconnected enough, the within-team trust collapses, and cooperation gives way to cheat-cheat equilibrium.

(ii) There are several different institutions that can eliminate the problem of free riding. One type are reputation aggregators that relax the link between the clustering coefficient and the formation of reputation. For instance, many internet-based apps face the problem of asymmetric information and cheating among clients, and solved it by some kind of collective reputation mechanism. The rating systems in E-bay, Uber, or Airbnb are good illustrations.

A second type of institutions eliminates the problem of strategic cooperation entirely. These alter the payoff matrices in a way that the Nash equilibrium shifts from the tragedy of the commons to the collective optimum. This can be done, for instance, by a collectively maintained institution that decreases the payoffs associated with the 'cheat' strategies enough to be dominated by the 'cooperate' strategies. Judicial systems are good examples for this, independent whether they are backed by official police, or enforced by an extra-legal system.

Notice that both kinds of institutions (reputation aggregators and payoff modifiers) are solutions to problems that tend to emerge when group size increases substantially. These, thus, allow the society to push towards a less integrated network structure without the society losing cooperativeness.

It is interesting to consider this result vis a vis two further observations. First, studies of innovation rates in urban spaces showed that high macro-level interconnectedness (and by implication low clustering coefficient) leads to higher innovation and technology dispersal rates (76). Second, humans have an (at least part) inherited drive towards engineering their personal social networks towards higher clustering coefficient pattern (as opposed to the meso-level structure). When people live in a social network with low interconnectedness, they feel socially depressed and lonely, even if the number social connection stay unchanged. Notice that from the combination of these two observations, the rise of different cooperativeness solutions (modern law, reputation accounting apps) should lead to more innovative but also more depressed and lonely societies.

(iii) Any kind of hierarchical structures of mass communication could be regarded as an example in which the synchronisation problem is eliminated. There are, however, two problems with these.

First, although humans tolerate mildly steep social hierarchies, in the 'ancestral' state it is most likely that our species lived in egalitarian social system. Most modern forager-hunter-gatherer cultures live today in levelled societies. And although there are exceptions to this general rule (e.g., North American fishing cultures, or Australian Native Aboriginal cultures), these are the minority, and most likely driven by an ecological logic. In any case, large societies with extremely steep hierarchies are rare in history, an even when they emerge, they either do not propagate through time, or results in a revolution. (There is no documented society in history or existing today in which the Gini coefficient went above 0.65, and most are well under this level.)

Second hierarchical systems are command systems, rather than synchronisation systems. In these, information tends to flow in one direction (e.g., in military structures and dictatorships), with the obvious drawback of losing out on the benefits of multi-way coordination, from integrating information from a wide range of origins.

Perhaps Wikipedia could serve as a non-hierarchical solution for en-mass information synchronisation, in which 'local' communities can be fully bonded and highly integrated, especially among editors creating reputation-based trust network.

### Consequence: meso-structures

As there is substantial variation in meso-structures of the graph library (Fig. 6), and these do matter for the trade-off between cooperation and coordination (Fig. 7), structural characteristics might emerge as optimal, e.g., within-graph hierarchical community structure might be beneficial for more efficient collaboration. Finding these will be the subject of future research.

Furthermore, previous research showed that inequality dynamics affects the network meso-structure in way that slows coordination dynamics (68), and – in retrospect – increases the clustering coefficient at the same time. This means that institutions regulating inequality might impact the behavioural synchrony-cooperation trade-off, further complicating the collaboration dilemma. This is especially so given that at least some aspects of status management are inherited behaviours (hierarchy awareness, self-status alert, third-party intervention, and, perhaps, alpha self-limiting). Thus, social technologies that regulate status hierarchy and social inequality do also interact with the social technologies that regulate the integratedness of the society, as well as the possible meso-structures discussed above. What we should see is a complex body of network regulating traditions, rules, and institutions, in which separating which social technology does what might be just as difficult as separating which set of genes is responsible for what behaviour.

# Supplementary Material

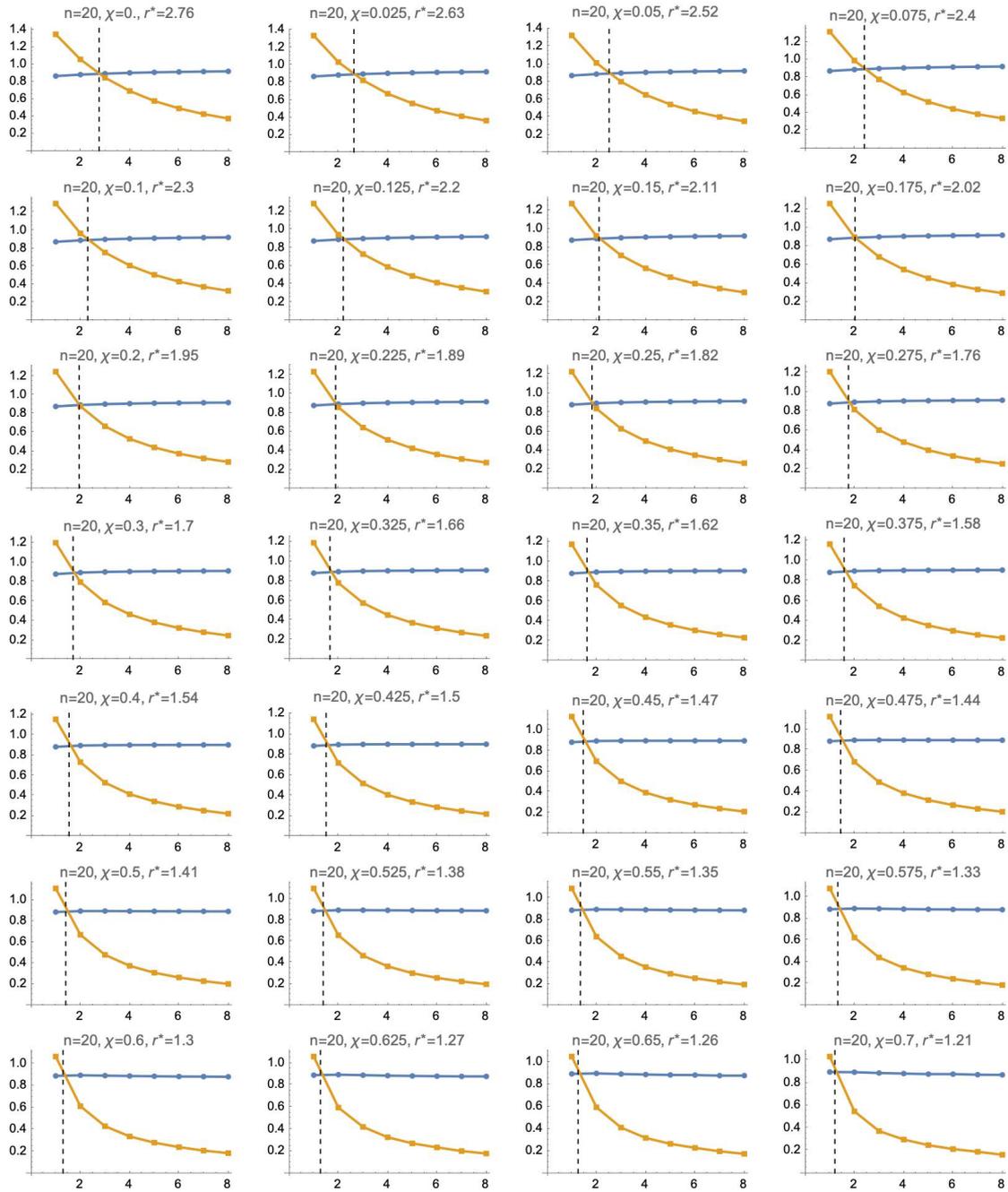

Fig. S1. Illustration for calculating the $r^*$ value, n=20.